\documentstyle[12pt,epsf,fleqn]{elsart}

\def \ni{\noindent}

\def \be {\begin{equation}}
\def \ee {\end{equation}}
\def \HM {HEIDEL\-BERG-MOSCOW~}

\begin{document}

\ni {\bf Corresponding author}\\
Prof. Dr. H.V. Klapdor-Kleingrothaus\\
Max-Planck-Institut f\"ur Kernphysik\\
Saupfercheckweg 1\\
D-69117 HEIDELBERG\\
GERMANY\\
Phone Office: +49-(0)6221-516-262\\
Fax: +49-(0)6221-516-540\\
email: klapdor@gustav.mpi-hd.mpg.de\\

\begin{frontmatter}
\title{Data Acquisition and Analysis of the $^{76}{Ge}$ 
        Double Beta Experiment in  Gran Sasso 1990-2003}

\author{H.V. Klapdor-Kleingrothaus}
\footnote{Spokesman of HEIDELBERG-MOSCOW (and GENIUS) Collaboration,\\
E-mail: klapdor@gustav.mpi-hd.mpg.de,\\
 Home-page: $http://www.mpi-hd.mpg.de.non\_acc/$}
{A. Dietz, I.V. Krivosheina}
\footnote{On leave of the Radiophysical-Research Institute, Nishnii-Novgorod, Russia}
\protect\newline {and O. Chkvorets} 

\address{Max-Planck-Institut f\"ur Kernphysik, PO 10 39 80,
  D-69029 Heidelberg, Germany}

\date{19.11.2003}

%%%%%%%%%%%%%%%%%%%%%%% abstract %%%%%%%%%%%%%%%%%%%%%%%
\begin{abstract}
        Data acquisition in a long running underground experiment 
        has its specific experimental challenges, 
        concerning data acquisition, 
        stability of the experiment and background reduction. 
        These problems are addressed here for the 
        \HM experiment, which collected data in the period August
        1990 - May 2003. 
        The measurement and the analysis of the data is presented. 
        The duty cycle of the experiment was $\sim$80\%, 
        the collected statistics is 71.7\,kg\,y. 
        The background achieved in the energy region 
        of the Q value for double beta decay 
        is 0.11\,events/\,kg\,y\,keV.
        The two-neutrino accompanied half-life is determined 
        on the basis of more than 
\protect\newline 100 000\,events. 
        The confidence level for the neutrinoless signal has been  
        improved to a 4$\sigma$ level.
\end{abstract}
\end{frontmatter}
%%%%%%%%%%%%%%%%%%%%%%% end abstract %%%%%%%%%%%%%%%%%%%%%%%

\section{Introduction}

%%%%%%%%%%%%%%%% new 07.10.2003 %%%%%%%%%%%%%%

        Since 40 years huge experimental efforts have gone 
        into the investigation of nuclear double beta decay 
        which probably is the most sensitive way to look for (total) 
        lepton number violation and probably the only way 
        to decide the Dirac or Majorana nature of the neutrino. 
        It has further perspectives 
        to probe also other types of beyond standard model physics. 
        This thorny way has been documented recently 
        in some detail 
%\cite{KK60Y,Kla95/98}.
[33,69].

        The half-lives to explore lying, with the order 
        of 10$^{25}$\,years, in a range  
        on 'half way' to that of proton decay, 
        the two main experimental problems were to achieve 
        a sufficient amount of double beta emitter material (source 
        strength) and to reduce the background in such experiment 
        to an extremely low level. 
        In both directions large progress has been 
        made over the decades. 
        While the first experiment using source as detector 
%\cite{Goldh66}, 
[59],
        had only grams of material to its disposal 
        (10.6\,g of CaF$_2$), in the last years up 
        to more than 10\,kg of enriched emitter material 
        have been used. 
        Simultaneously the background of the experiments 
        has been reduced strongly over the last 40\,years. 
        E.g. compared to the  first Germanium $\beta\beta$ experiment 
%\cite{1DBD-exp}, 
[81],   
        working still with natural Germanium, containing 
        the double beta emitter $^{76}{Ge}$ only with 7.8\%, 
        40\,years later 
        the background in the \HM experiment 
        is reduced by a factor of 10$^4$.

        Nevertheless experiments have to run over years 
        to collect sufficient statistics and this led 
        to other experimental challenges: stable data acquisition 
        and calibration over long time periods (more than a decade 
        in the \HM experiment).
        The final dream behind all these efforts was less 
        to see a standard-model allowed second-order effect 
        of the weak interaction in the nucleus - the 
        two-neutrino-accompanied decay mode -  
        which has been observed meanwhile for about 10 nuclei - 
        to observe neutrinoless double beta decay, and with
        this a first hint of beyond standard model physics, 
        yielding at the same time a solution 
        of the absolute scale of the neutrino mass spectrum. 

        In this paper we describe how these challenges 
        have been mastered in the \HM experiment, 
        which is running in the Gran Sasso Underground Laboratory
        since August 1990, and which is now, 
        together with the R. Davis $^{37}{Cl}$ solar neutrino experiment 
%\cite{Davis},
[61], 
        the Baksan Neutrino Scintillation telescope 
%\cite{IK-SN87}
[62]
        which saw together with Kamiokande 
        and IMB the first Supernovae neutrinos 
%\cite{KAM87,IMB87},
[63,64], 
        Kamiokande and SuperKamiokande 
%\cite{Kamiok}, 
[65],
        Gallex (GNO) 
%\cite{Gal}
[66] 
        and SAGE 
%\cite{Gal},
[66],
        and DAMA 
%\cite{DAMA-03},
[104],
        one of the long-running underground experiments.
        
        In our last publications 
%\cite{KK02-Found,KK02,KK02-PN}
[3,1,2]
        we presented an analysis of the data taken until May 2000. 
        Since then we have taken three more years of data.  
        In this paper we report the result of the measurements 
        over the full period August 1990 until May 2003. 
        The better quality of the new data and of the present 
        analysis, which has been improved in various respects, 
        allowed us to significantly improve the investigation 
        of the neutrinoless double beta decay process, 
        and to deduce more stringent values of its parameters.  

%%%%%%%%%%%%%%%% end new 07.10.2003 %%%%%%%%%%%%%%

%%%%%%%%%%%%%%%% Section 2 %%%%%%%%%%%%%%%%%%%%%
\section{Performance of the Experiment and Data Taking}
%%%%%%%%%%%%%%%% Section 2 %%%%%%%%%%%%%%%%%%%%%

\Huge

{\it  Comment:}

        {\bf full version \\
         (50 pages, 36 figures, 9 tables), \\
        zipped ps file (size $\sim$ 21\,MB)\\ 
        or pdf format (size  $\sim$ 26\,MB), \\
        can be found in \\
         http://www.mpi-hd.mpg.de/\\
        non$\_$acc/super.html 

        In press in NIM A, 2004.}

\end{document}